\documentclass[fp,twocolumn,dvipdfmx]{jpsj3}
\usepackage[dvipdfmx]{graphicx}
\def\S{Sect. }

\def\d{{\rm d}}
\def\q{{\rm q}}

\def\msp{m_{\rm sp}}
\def\v#1{\mib #1}

\def\H{{\mathcal H}}

\newcommand{\aver}[1]{\left\langle {#1} \right\rangle}
\title
{
Infinite Series of Ferrimagnetic Phases Emergent from the Gapless Spin Liquid Phase of Mixed Diamond Chains
}

\author
{
Kazuo Hida\thanks{E-mail address: hida@mail.saitama-u.ac.jp}
}

\inst
{Professor Emeritus, Division of Material Science, Graduate School of Science and Engineering, \\ 
Saitama University, Saitama, Saitama, 338-8570\\ 
}

\recdate
{}

\abst
{
The ground-state phases of mixed diamond chains with ($S, \tau^{(1)}, \tau^{(2)})=(1/2,1/2,1)$, where $S$ is the magnitude of vertex spins, and $\tau^{(1)}$ and $\tau^{(2)}$ are those of apical spins, are investigated. 
  The apical spins $\v{\tau}^{(1)}$ and $\v{\tau}^{(2)}$ are connected with each other by an exchange coupling $\lambda$. Other exchange couplings are set equal to unity. This model has an infinite number of local conservation laws. For large $\lambda$, the ground state is equivalent to that of the uniform spin $1/2$ chain. Hence, the ground state is a gapless spin liquid. For  $\lambda \leq 0$, the ground state is a Lieb-Mattis ferrimagnetic phase with spontaneous magnetization $\msp=1$ per unit cell.  For intermediate $\lambda$, we find a series of ferrimagnetic phases with $\msp=1/p$ where $p$ takes positive integer values.  The phases with $p \geq 2$ are accompanied by the spontaneous breakdown of the $p$-fold translational symmetry. It is suggested that the phase with arbitrarily large $p$, namely infinitesimal spontaneous magnetization, is allowed as $\lambda$ approaches the transition point to the gapless spin liquid phase.
}
\begin{document}
\maketitle
\section{Introduction}

The quantum effects in low-dimensional frustrated magnets have been extensively studied in recent condensed matter physics.\cite{intfrust,diep} Various exotic quantum phases emerge from the interplay of quantum fluctuation and frustration. 
Among them, the diamond chain\cite{Takano-K-S,ht2017,kiku2,kiku3,hida2019,hts_dist_mdc,hida2020}, whose lattice structure is shown in Fig. \ref{lattice},  is known as a model with an infinite number of local conservation laws. The ground states can be classified by the corresponding quantum numbers. If the two apical spins have equal magnitudes, which is the case widely investigated, each pair of apical spins can form a singlet dimer. It cuts the correlation between both sides and the ground state is a direct product of the cluster ground states separated by dimers.

The ground states of spin-1/2 diamond chains have been investigated in Ref. \citen{Takano-K-S}. In addition to the spin cluster ground states, the ferrimagnetic state with spontaneous magnetization $\msp=1/2$ per unit cell is found. In the latter phase, the apical spins form triplet dimers and all the spins collectively form a long-range ordered ferrimagnetic state.  
Extensive experimental studies have been also carried out on the magnetic properties of natural mineral azurite that is regarded as an example of distorted spin-1/2 diamond chains.\cite{kiku2,kiku3}

The ground states of spin-1 diamond chains have been also investigated in Refs. \citen{Takano-K-S} and \citen{ht2017}. In addition to the spin cluster ground states, the nonmagnetic Haldane state and the ferrimagnetic states with spontaneous magnetization $\msp=1$ and 1/2 are found. It should be noted that the latter ferrimagnetic state is accompanied by  a spontaneous translational symmetry breakdown. 
 
On the other hand, if the magnitudes of the two apical spins are not equal to each other, they cannot form a singlet dimer. Hence, all spins in the chain inevitably form a many-body correlated state. In many cases,  (quasi-)long-range order evolves including the vertex spins. As a simple example of such cases, we investigate the mixed diamond chain with apical spins of magnitude 1 and 1/2, and vertex spins, 1/2 in the present work. Remarkably, we find an infinite series of ferrimagnetic phases.

This paper is organized as follows. 
 {In \S 2}, the model Hamiltonian is presented. 
 {In \S 3}, the ground-state phase diagram is determined numerically. The behavior of the spontaneous magnetization in each phase is presented and analyzed. 
The last section is devoted to a summary and discussion.

\section{Hamiltonian}

\begin{figure}[t] 
\centerline{\includegraphics[width=7cm]{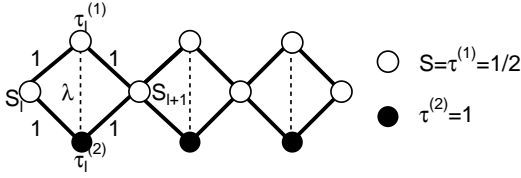}}
\caption{Structure of the diamond chain investigated in this work  $S=\tau^{(1)}=1/2$ and $\tau^{(2)}=1$.}
\label{lattice}
\end{figure}

We consider the Hamiltonian 
\begin{align}
{\mathcal H} &= \sum_{l=1}^{L} \left[\v{S}_{l}(\v{\tau}^{(1)}_{l}+\v{\tau}^{(2)}_{l}) \right. 
\nonumber\\
&+(\v{\tau}^{(1)}_{l}+\v{\tau}^{(2)}_{l})\v{S}_{l+1}
+ \left. \lambda\v{\tau}^{(1)}_{l}\v{\tau}^{(2)}_{l}\right] , 
\label{hama}
\end{align}
where $\v{S}_{l}, \v{\tau}^{(1)}_{l}$ and $\v{\tau}^{(2)}_{l}$ are spin operators with magnitudes ${S}_{l}={\tau}^{(1)}_{l}=1/2$ and ${\tau}^{(2)}_{l}=1$. The number of the unit cells is denoted by $L$, and the total number of sites is $3L$. Here, the parameter $\lambda$ controls the frustration as depicted in Fig. \ref{lattice}. 

The Hamiltonian (\ref{hama}) has a series of local conservation laws. 
To see it, we rewrite Eq. (\ref{hama}) 
 in the form, 
\begin{align}
\H &= \sum_{l=1}^{L} \left[\v{S}_{l}\v{T}_{l}+\v{T}_{l}\v{S}_{l+1}
+ \frac{\lambda}{2}\left(\v{T}^2_{l}-\frac{11}{4}\right)\right],
\label{ham2} 
\end{align}
where the composite spin operators $\v{T}_l$ are defined as 
\begin{align}
\v{T}_{l} \equiv \v{\tau}^{(1)}_{l}+\v{\tau}^{(2)}_{l} 
\quad (l = 1, 2, \cdots ,L). 
\end{align}
Then, it is evident that 
\begin{align}
[\v{T}_l^2, {\mathcal H}] = 0 \quad (l = 1, 2, \cdots , L). \label{eq:cons}
\end{align}
Thus, we have $L$ conserved quantities $\v{T}_l^2$ for all $l$. 
By defining the magnitude $T_l$ of the composite spin $\v{T}_l$ by $\v{T}_l^2 = T_l (T_l + 1)$, we have a 
 set of good quantum numbers $\{T_l; l=1,2,...L\}$ where $T_l=$ 1/2 and 3/2. 
The total Hilbert space of the Hamiltonian (\ref{ham2}) consists of 
separated subspaces, each of which is specified by 
a definite set of $\{T_l\}$, i.e., a sequence of 1/2 and 3/2. 
 A pair of apical spins with $T_l=1/2$ is called a doublet (hereafter abbreviated as d) and that with $T_l=3/2$ a quartet (abbreviated as q). 

\section{Ground-State Phase Diagram}

\subsection{Ground states for $\lambda \gg 1$}\label{sect:nonmag}

For $\lambda \gg 1$, $\forall l \ T_l=1/2$. Hence, this model is equivalent to the spin-1/2 antiferromagnetic Heisenberg chain whose ground state is a gapless spin liquid. 
\subsection{Ground states for $\lambda \ll 1$}\label{sect:full}

For $\lambda \ll 1$, $\forall l \ T_l=3/2$. Hence, this model is equivalent to the spin-1/2-3/2 alternating antiferromagnetic Heisenberg chain whose ground state is a ferrimagnetic state with spontaneous magnetization $\msp=1$ per unit cell according to the Lieb-Mattis theorem.\cite{Lieb-Mattis} Here, $\msp$ is defined by
\begin{align}
\msp=\frac{1}{L}\sum_{l=1}^L(\aver{S^z_l}+\aver{T^z_l})
\end{align}
where $\aver{}$ denotes the expectation value in the ground state with an infinitesimal symmetry breaking magnetic field in $z$-direction. 
\subsection{Intermediate $\lambda$}

\begin{figure} 
\centerline{\includegraphics[width=5cm]{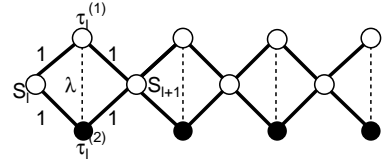}}
\caption{Lattice structure used for the finite size DMRG calculation}
\label{fig:lattice_finite}
\end{figure}

In the absence of spontaneous translational symmetry breakdown, only above two phases are allowed. To pursue the possibility of other phases, we employ the finite size DMRG method with the geometry of Fig. \ref{fig:lattice_finite}. The corresponding Hamiltonian is given by
\begin{align}
\H &= \sum_{l=1}^{L}\v{S}_{l}\v{T}_{l}+ \sum_{l=1}^{L-1}\v{T}_{l}\v{S}_{l+1}
+\sum_{l=1}^{L} \frac{\lambda}{2}\left(\v{T}^2_{l}-\frac{11}{4}\right). \label{ham3}
\end{align}
This geometry is chosen to allow for the nonmagnetic ground state for $\lambda \gg 1$. Here and in what follows,  the number of states $\chi$ kept in each subsystem in the DMRG calculation ranged from 240 to 360. 
We find that the results with $\chi=240$ are accurate enough in the present work. 

The ground-state energies of the Hamiltonian (\ref{ham3}) up to $L=16$ for all possible configurations $\{T_l\}$ are calculated. The configurations that give the lowest energy are identified for each $\lambda$.  
The spontaneous magnetization per unit cell is given by
\begin{align}
\msp=\frac{1}{L}\left|\sum_{l=1}^{L}(T_l-S_l)\right|
\end{align}
from the Lieb-Mattis theorem\cite{Lieb-Mattis} for the Hamiltonian (\ref{ham3}).

\begin{figure} 
\centerline{\includegraphics[width=7cm]{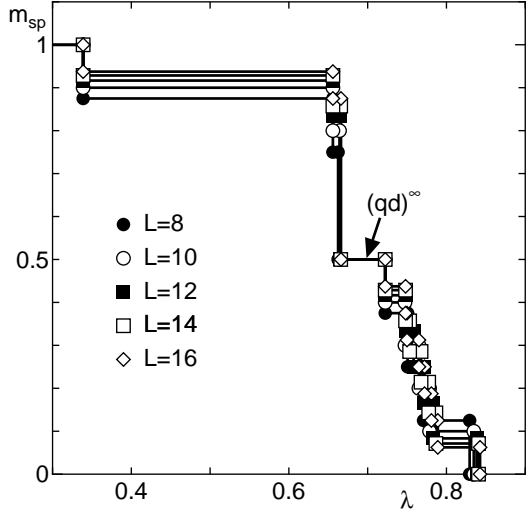}}
\caption{$\lambda$-dependence of $\msp$ calculated by the finite size DMRG method.}
\label{fig:mag_finite}
\end{figure}

Figure \ref{fig:mag_finite} shows our numerical result for the $\lambda$-dependence of $\msp$. 
In addition to the two phases described in Sect. \ref{sect:nonmag} and Sect \ref{sect:full}, a quantized ferrimagnetic ground state with $\msp=1/2$ is clearly observed. This state has the configuration with $T_l=\q$ for odd $l$ and $T_l=\d$ for even $l$. We denote this configuration as $(\q\d)^{L/2}$.  In the thermodynamic limit, this configuration tends to the $(\q\d)^{\infty}$ configuration. This configuration is equivalent to the $(\d\q)^{\infty}$ configuration in the thermodynamic limit. Hence, this phase is doubly degenerate and is accompanied by the spontaneous breakdown of twofold translational symmetry. In what follows, similar notations are employed for other configurations. Within the finite-size DMRG calculation, the intermediate ferrimagnetic phases are observed between this phase and the nonmagnetic phase. Although two intermediate ferrimagnetic phases are also observed near $\msp=1$,  these are artifacts of the finite size effect, since they correspond to $\msp=1-1/L$ and $1-2/L$.

For finite chains, however, the values of $\msp$ are constrained by the system size. Hence, it is not clear whether $\msp$ is quantized or continuously varying with $\lambda$ for $0 < \msp < 1/2$ in the thermodynamic limit.

\begin{figure} 
\centerline{\includegraphics[width=7cm]{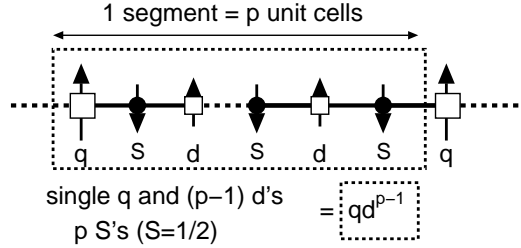}}
\caption{$(\q\d^{p-1})^{\infty}$ configuration.}
\label{fig:periodic_phase}
\end{figure}

Hence, we resort to the infinite-size DMRG calculation. However, it is not possible to carry out the calculation for all possible configurations of $\{T_l\}$ for infinite chains. Hence, we analyze this regime in the following way: We start with plausible candidates of periodic configurations of $T_l$'s and find the configuration that gives the lowest energy ground state among them. This leads to a stepwise $\lambda$-dependence of $\msp$. Then, we check the stability of these steps against the formation of defects. As plausible candidates of the ground states, we consider the configurations $(\q\d^{p-1})^{\infty}$ with $\msp=1/p$ that consist of an infinite array of segments $\q\d^{p-1}$ with length of $p$ unit cells as depicted in Fig. \ref{fig:periodic_phase}. These configurations can be obtained from the configuration $\d^{\infty}$  corresponding to the nonmagnetic ground state by inserting a $\q$ every $p$ unit cells periodically. It should be remarked that these states are accompanied by the $p$-fold spontaneous translational symmetry breakdown. Hereafter, this series of configurations is called the main series configurations. The case $p=1$ corresponds to the $\q^{\infty}$ configuration that corresponds to the Lieb-Mattis type ferrimagnetic phase with $\msp=1$.

\begin{figure} 
\centerline{\includegraphics[width=7cm]{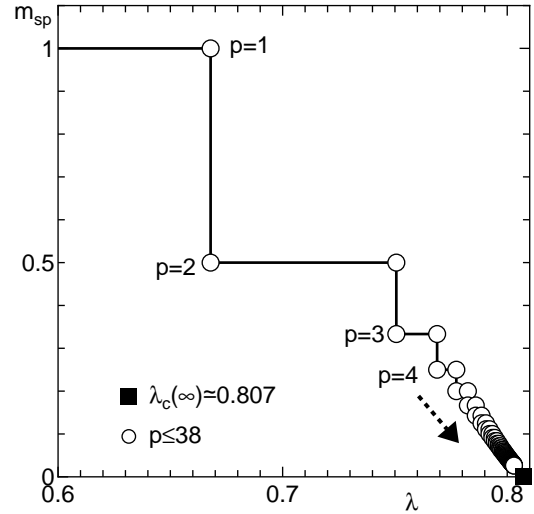}}
\caption{$\lambda$-dependence of $\msp$ for the main series configurations calculated by the infinite size DMRG method.}
\label{fig:mag_infinite}
\end{figure}

The $\lambda$-dependence of $\msp$ in the main series is shown in Fig. \ref{fig:mag_infinite}. 
\begin{figure} 
\centerline{\includegraphics[width=7cm]{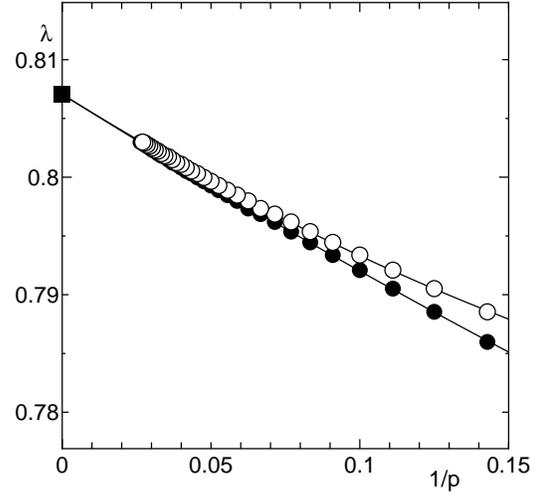}}
\caption{Extrapolation scheme of $\lambda_{\rm c}(p,p-1)$ and  $\lambda_{\rm c}(p+1,p)$ to $p \rightarrow \infty$. The filled and open circles correspond to the fits by Eq. (\ref{eq:lamc1}) and Eq. (\ref{eq:lamc2}), respectively. The filled square is the extrapolated value $\lambda_{\rm c}(\infty)$.}
\label{fig:extra}
\end{figure}
\begin{figure} 
\centerline{\includegraphics[width=7cm]{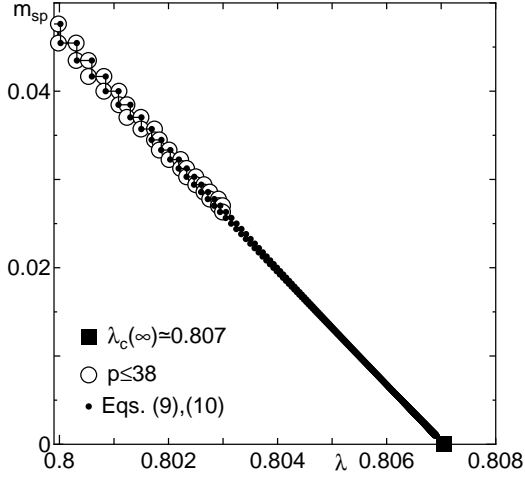}}
\caption{$\lambda$-dependence of $\msp$ for the main series configurations near $\lambda=\lambda_{\rm c}(\infty)$. The open circles are the results of the infinite size DMRG calculation. The small filled circles and solid lines are extrapolation by Eqs. (\ref{eq:lamc1}) and (\ref{eq:lamc2}).}
\label{fig:mag_infinite_ext}
\end{figure}
The spontaneous magnetization $\msp$ rises continuously from $\msp=0$ 
at the critical value of $\lambda$ given by
\begin{align}
\lambda_{\rm c}(\infty)\equiv\lim_{p\rightarrow \infty}\lambda_{\rm c}(p,p-1)\simeq 0.807
\end{align}
where the boundary between the  $(\q\d^{p-1})^{\infty}$ phase with $\msp=1/p$ and $(\q\d^{p-2})^{\infty}$ phase with $\msp=1/(p-1)$ is denoted by $\lambda_{\rm c}(p,p-1)$. The extrapolation to $p \rightarrow \infty$ is carried out using the data for $38 \geq p\geq 11$ assuming the following two asymptotic forms
\begin{align}
\lambda_{\rm c}(p,p-1)&=\lambda_{\rm c}(\infty)+\frac{C_1}{p}+\frac{C_2}{p^2},\label{eq:lamc1}\\
\lambda_{\rm c}(p+1,p)&=\lambda_{\rm c}(\infty)+\frac{C'_1}{p}+\frac{C'_2}{p^2}.\label{eq:lamc2}
\end{align}
The extrapolation procedure is plotted in Fig. \ref{fig:extra}. 
The both extrapolations give the same value for $\lambda_{\rm c}(\infty)$ up to the above digit.  It should be noted that Eq. (\ref{eq:lamc1}) and  Eq. (\ref{eq:lamc2}) correspond to the extrapolation of left and right ends of the steps, respectively. Using Eqs. (\ref{eq:lamc1}) and  (\ref{eq:lamc2}), the $\lambda$-dependence of $\msp$ is plotted down to $\msp=0$ in Fig. \ref{fig:mag_infinite_ext}.

The remaining question is whether the intermediate configurations with $1/p <\msp <1/(p-1)$ can be a ground state. To answer this question, it is necessary to calculate the ground-state energies for all possible configurations of $\{T_l\}$, which is impossible. Hence, we confine ourselves to the following configurations that are plausible to compete with the $(\q\d^{p-1})^{\infty}$ configurations.

\begin{enumerate}

\begin{figure} 
\centerline{\includegraphics[width=8cm]{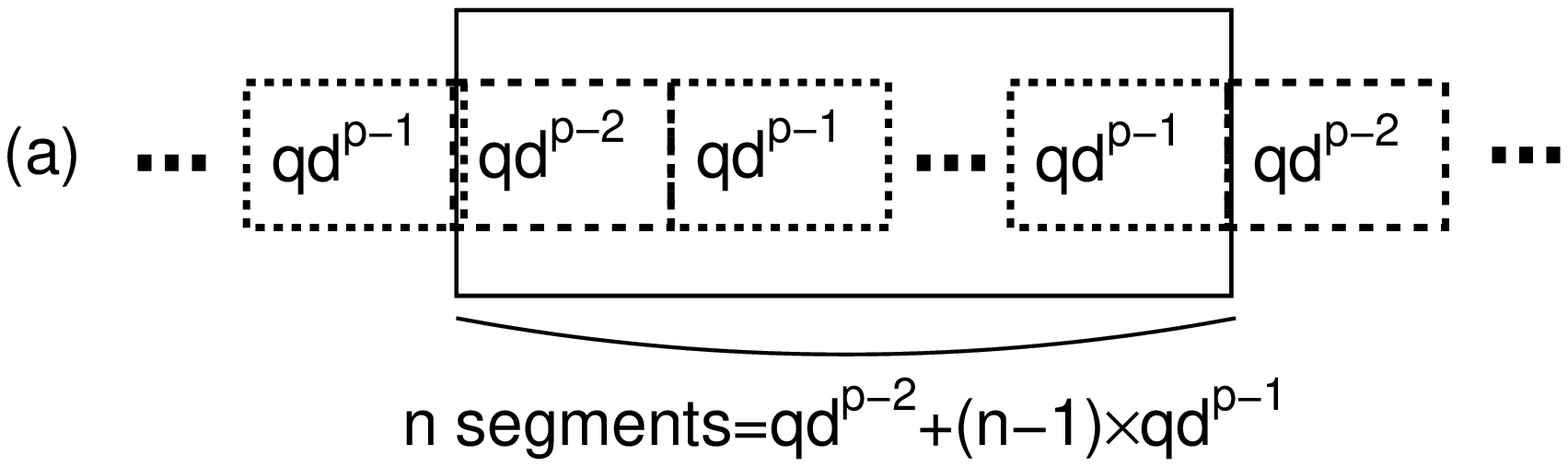}}
\vspace{4mm}
\centerline{\includegraphics[width=8cm]{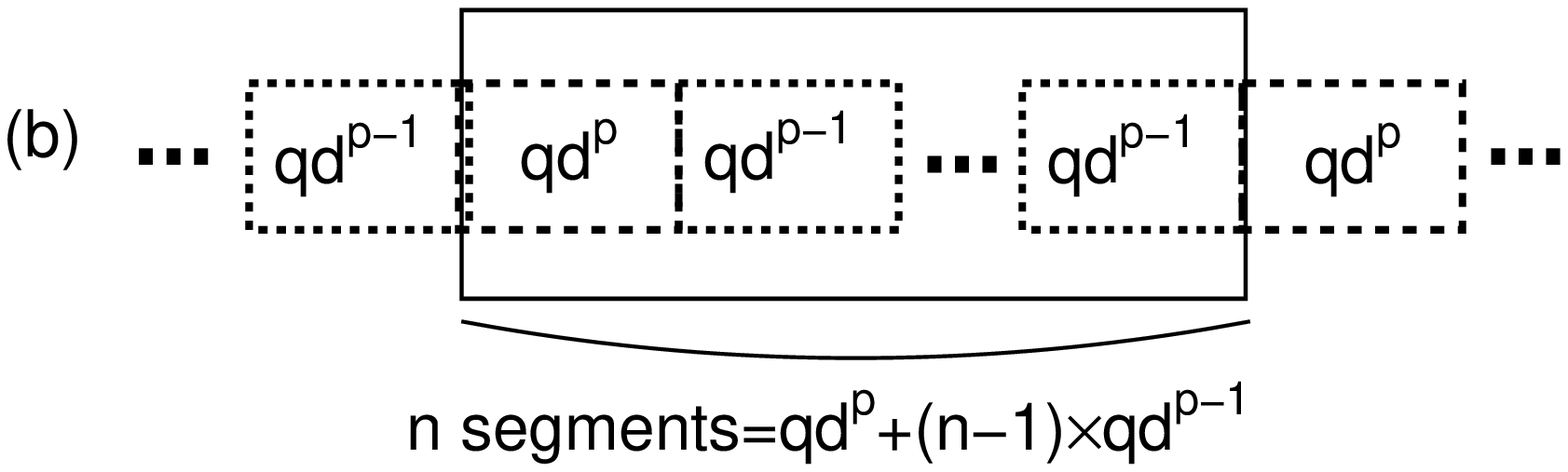}}
\caption{(a) $(\q\d^{p-2}(\q\d^{p-1})^{n-1})^{\infty}$  configuration and (b) $\q\d^p((\q\d^{p-1})^{n-1})^{\infty}$ configuration.}
\label{fig:defect_phase}
\end{figure}

\item $(\q\d^{p-2}(\q\d^{p-1})^{n-1})^{\infty}$ configuration ($\msp=n/(p(n-1)+(p-1))$) depicted in Fig. \ref{fig:defect_phase}(a): A $\q\d^{p-1}$ segment is replaced by  a $\q\d^{p-2}$ segment per every $n$ 
 segments in the $(\q\d^{p-1})^{\infty}$ configuration. For $n=1$, this configuration reduces to the $(\q\d^{p-2})^{\infty}$ configuration with $\msp={1}/{(p-1)}$ that corresponds to the neighboring step  of the main series with higher $\msp$. For $n=2$, this configuration reduces to that with alternating  $\q\d^{p-1}$ and $\q\d^{p-2}$ segments.

\item $(\q\d^p(\q\d^{p-1})^{n-1})^{\infty}$ configuration ($\msp={n}/({p(n-1)+(p+1)})$) depicted in Fig. \ref{fig:defect_phase}(b): A $\q\d^{p-1}$ segment is replaced by  a $\q\d^p$ segment per every $n$ 
segments in the $(\q\d^{p-1})^{\infty}$ configuration. For $n=1$, this configuration reduces to the $(\q\d^p)^{\infty}$ configuration with $\msp={1}/{(p+1)}$ that corresponds to the neighboring step  of the main series with lower $\msp$. For $n=2$, this configuration reduces to that with alternating  $\q\d^{p-1}$ and $\q\d^p$ segments. 

\item The configurations with spatial periodicity less than 12 and spontaneous magnetization $\msp \leq 1/4$ have been thoroughly checked.
\end{enumerate}

 For configurations (1) and (2), the computational cost increases with an increase of $p$ and $n$. We have numerically confirmed that these states with $2 \leq n \leq 4$ are not the ground state for  $2 \leq p \leq 11$. It is further confirmed that those  with  $2 \leq n \leq 6$ are not the ground state for $2 \leq p \leq 8$. Within our numerical accuracy, we find no configurations that give lower energy than the main series configurations. Within the available numerical data, the configurations with higher $n$ are even less favorable. Hence, it is highly plausible that the state intermediate $\msp$ is not a ground state at least for $p \le 11$. Also, configurations (3) do not give the ground state except for the main series configurations. Thus, we expect that only the configurations in the main series are realized in the ground state.

\section{Summary and Discussion}

 The ground-state phases of diamond chains (\ref{hama}) with $(S,\tau^{(1)},\tau^{(2)})=(1/2,1/2,1)$  are investigated. Between the gapless spin liquid phase for large $\lambda$ and Lieb-Mattis ferrimagnetic phase with $\msp=1$ for $\lambda \leq 0$, we find a series of quantized ferrimagnetic phases with $\msp=1/p$ where $p$ takes all positive integer values. 

The $\lambda$-dependence of the spontaneous magnetization $\msp$ is very different from other diamond chains with ferrimagnetic ground states. Although the quantized ferrimagnetic phases are present even in undistorted diamond chains, the allowed values of spontaneous magnetization are limited to several simple rational values.\cite{Takano-K-S,ht2017}

There are some examples of ferrimagnetic ground states of diamond chains that are induced by the lattice distortion.\cite{hida2019,hts_dist_mdc} In these cases, the ground state of the undistorted chain is a paramagnetic state consisting of clusters with finite magnetic moments. The sizes of the clusters are limited even in the absence of distortion. The distortions induce ferromagnetic interactions between the cluster spins leading to the quantized ferrimagnetic phases. The quantum fluctuations of the lengths of clusters are also induced by distortion leading to the partial ferrimagnetic phases.\cite{hts_dist_mdc}

In the present case, the ferrimagnetic phases are present even in the absence of distortion. In contrast to the cases of Refs. \citen{Takano-K-S,ht2017,hida2019,hts_dist_mdc}, arbitrarily large segments are allowed leading to the infinitesimally small steps around $\lambda =\lambda_{\rm c}(\infty)$. However, no fluctuations of the lengths of the segments are allowed in the present model, since the magnitudes of the composite spins $T_l$ remain good quantum numbers. Hence, partial ferrimagnetic phases are absent in contrast to the cases of Refs. \citen{hts_dist_mdc} and \citen{hida2019}. 

In the spin-1 alternating bond diamond chain with bond alternation $\delta$, the nonmagnetic phase is equivalent to the ground state of the spin-1 alternating bond Heisenberg chain with bond alternation $\delta$.\cite{hida2020} In this model, an intermediate ferrimagnetic phase is observed in the tiny region close neighborhood of the point  $(\lambda,\delta)=(\lambda_{\rm c},\delta_{\rm c})\simeq (1.0832,0.2598)$ that  corresponds to the endpoint of the Haldane-dimer critical line.\cite{Kato-Tanaka1994,Yamamoto1994,Totsuka1995,Kitazawa-Nomura1997}.  In Ref. \citen{hida2020}, it has been speculated that this region is the partial ferrimagnetic phase. However, considering the similarity of the gapless spin-liquid phase of the present model and the Haldane-dimer critical line of the  spin-1 alternating bond diamond chain, it would be more reasonable to speculate that the infinite series of quantized ferrimagnetic phases similar to those discussed in the present work is realized also in this case. Unfortunately, the numerical confirmation is difficult due to the smallness of the width of this region. 

As mentioned above, in some examples of the quantized ferrimagnetic phases in undistorted diamond chains, the allowed values of spontaneous magnetization are limited to several rational values.\cite{Takano-K-S,ht2017} In these examples, the nonmagnetic phases neighboring the ferrimagnetic phases are spin-gap phases. On the other hand, the nonmagnetic phase neighboring the ferrimagnetic phase in the present model with infinitesimal step is the gapless spin liquid phase. This seems to suggest that the infinitesimal energy scale of the gapless spin liquid phase helps the emergence of the exotic ferrimagnetic phase with infinitesimal spontaneous magnetization. A further analytical approach would be required to get insight into the physical implication of the present phenomenon. 

So far, the infinite series of ferrimagnetic phases proposed in this work have not been found in real materials. However, since the gapless spin liquid phases are generic critical states in quantum spin chains, it would be possible that these series of phases are realized in the presence of appropriate frustrating exchange interactions. Nevertheless, in more realistic cases, the perturbation that does not preserve the conservation laws (\ref{eq:cons}) is inevitable. In such cases, the infinitesimal structure of spontaneous magnetization might be smeared. In this context, it would be an interesting problem to investigate the effect of lattice distortion in the present model. These studies are left for future investigation.
\acknowledgments

A part of the numerical computation in this work has been carried out using the facilities of the Supercomputer Center, Institute for Solid State Physics, University of Tokyo, and Yukawa Institute Computer Facility at Kyoto University.

\end{document}